\documentclass[aps,pra,twocolumn,superscriptaddress]{revtex4}

\usepackage{amsmath,amsfonts,amssymb}
\usepackage{graphicx,float,calc}
\usepackage{color,bm}
\usepackage{ulem}
\usepackage{braket}
\usepackage{hyperref}
\usepackage{graphicx,epstopdf}
\usepackage{CJK}
\usepackage{makecell}
\usepackage{multirow}
\usepackage{array}

\newcommand{\ehbar}{\hbar_{\text{eff}}}

\begin{document}

\title{Phase modulation of directed transport, energy diffusion and quantum scrambling in a Floquet non-Hermitian system}

\author{Wen-Lei Zhao}
\email[]{wlzhao@jxust.edu.cn}
\affiliation{School of Science, Jiangxi University of Science and Technology, Ganzhou 341000, China}

\author{Guanling Li}
\affiliation{School of Science, Jiangxi University of Science and Technology, Ganzhou 341000, China}

\author{Jie Liu}
\email[]{jliu@gscaep.ac.cn}
\affiliation{Graduate School of China Academy of Engineering Physics, Beijing 100193, China}
\affiliation{CAPT, HEDPS, and IFSA Collaborative Innovation Center of the Ministry of Education, Peking University, Beijing 100871, China}

\begin{abstract}
We investigate both theoretically and numerically the wavepacket's dynamics in momentum space for a Floquet non-Hermitian system with a periodically-kicked driven potential. We have deduced the exact expression of a time-evolving wavepacket under the condition of quantum resonance. With this analytical expression, we can investigate thoroughly the temporal behaviors of the directed transport, energy diffusion and quantum scrambling. We find interestingly that, by tuning the relative phase between the real part and imaginary part of the kicking potential, one can manipulate the directed propagation, energy diffusion and quantum scrambling efficiently: when the phase equals to $\pi/2$, we observe a maximum directed current and energy diffusion, while a minimum scrambling phenomenon protected by the $\mathcal{PT}$-symmetry; when the phase is $\pi$, both the directed transport and the energy diffusion are suppressed, in contrast, the quantum scrambling is enhanced by the non-Hermiticity. Possible applications of our findings are discussed.
\end{abstract}

\date{\today}

\maketitle

\setlength{\parindent}{2em}
\section{Introduction}

Engineering the wavepacket's dynamics, such as directed transport~\cite{Kenfack08prl,Hainaut18pra,ZQLi23pra,Poletti07pra}, energy diffusion~\cite{JBGong01prl,Bitter17prl,Sentef20prr,Bai20pra,Downing23} and information scrambling~\cite{Meier19pra,JHWang22prr,JHarris22prl}, is of great interest both theoretically and experimentally across various fields of physics~\cite{XWang23prl,Cheng19prl}.
The phase incorporated in Floquet driven potentials is vitally a knob to manipulate the quantum dynamics~\cite{JBGong02prl}. For example, the temporal modulation of the phase of laser standing waves can be used to create artificial gauge fields for ultracold neutral atoms, mimicking the transport behavior of electrons in a synthetic nanotube, with the Aharonov-Bohm flux controllable by the phase~\cite{Hainaut18nc,Garreau20pra,CTian10NJP}. The quasiperiodically modulated phase of the external driven potential even induces the formation of synthetic dimension~\cite{Casati89prl,Shepelyansky87,Tian11prl,Hainaut18prl}, wherein the Anderson metal-insulator transition of disorder systems is experimentally observed by using a variant of the kicked rotor model~\cite{Lopez13,Manai15prl,Lopez12prl}. More significantly, complex potentials are achievable in atom-optical experiments, where precise control over the relative phase between the real and imaginary components of this non-Hermitian potential allows for the realization of distinct symmetry classes of systems~\cite{Chudesnikov91,Keller97}.

Nowadays, non-Hermiticity is widely acknowledged as a fundamental extension of conventional quantum mechanics~\cite{Bender98prl,JBGong13Jpa} due to its natural incorporation of the gain and loss or non-reciprocity in diverse systems like photonic crystals~\cite{CWang21Sci,Ozdemir19,YSong19,HCLi19PB,YShen22PRAL,HCLi22Nano}, acoustic arrays~\cite{LZhang21nc,XZhu14prx}, and electrical  circuits~\cite{HXZhang23prb,Chitsazi17prl}.
The emergence of the complex eigenspectra of non-Hermitian systems gives rise to rich phenomena with no Hermitian counterpart. For example, as the system adiabatically crosses the exceptional points at which both the eigenvalues and eigenstates coalesce, the Landau-Zener tunneling emerges, indicating the breakdown of adiabaticity~\cite{KSim23prl,WYWang22pra,XQTong23prb,XWang23njp}. The spontaneous $\cal{PT}$-symmetry breaking induces the quantized acceleration of directed transport~\cite{WLZhao19pre} and the quantized response of quantum scrambling~\cite{Wlzhao22prr} in non-Hermitian chaotic systems~\cite{Wlzhao23pra}. In addition, various topological symmetry classes, including point gap and line gap eigenbands in the complex plane, have been recognized as having significant impacts on edge-state transport behavior, for instance non-Hermitian skin effects~\cite{Yokomizo19prl,FSong19prl,LWZhou23arx}. Besides, the non-Hermitian potential can be engineered effectively in versatile platforms of photonic systems~\cite{Meng23arx,HCao15rmp} and atom-optics~\cite{Keller97}, which unveils the possibilities for manipulating the wavepacket's dynamics in a controllable manner.

In this context, we investigate both theoretically and numerically the phase modulation of the directed transport, energy diffusion, and quantum scrambling, in a non-Hermitian quantum kicked rotor (NQKR) model with quantum resonance condition. We find that the relative phase between the real part and imaginary part of the kicking potential dominates the wavepacket's dynamics. Specifically, as time evolves, the direct current undergoes a transition from quadratic growth, dependent on the non-Hermitian driving strength, to linear growth, which becomes independent of the non-Hermitian driving strength. Such a dynamical crossover from non-Hermiticity dependent regime to non-Hermiticity independent regime is also observed in the ballistic diffusion of mean energy. Furthermore, we unveil the quadratic growth in the OTOCs during the initial stages of time evolution and a transition to linear growth after a sufficiently long period, both influenced by the non-Hermitian driving strength. Interestingly, when the phase equals to $\pi/2$ ensuring $\mathcal{PT}$-symmetry, we observe a maximum transport behavior and a minimum scrambling phenomenon. At a phase of $\pi$, both the directed transport and the energy diffusion are suppressed, in contrast the quantum scrambling is enhanced by the non-Hermiticity. Our findings provide theoretical guidance for Floquet engineering of quantum dynamics in non-Hermitian systems, which has significant implications in various physics fields, including quantum chaotic control and condensed matter physics.

The paper is organized as follows. In Sec.~\ref{SecMResl}
we describe the system and show the phase modulation of wavepacket's dynamics with emphasising on the directed current, energy diffusion and quantum scrambling. A summary is presented in Sec.~\ref{SecSum}.

\section{NQKR Model and main results}\label{SecMResl}
The dimensionless Hamiltonian of the NQKR reads
\begin{equation}\label{Hamil}
{\rm H}=\frac{{p}^2}{2}+ V_K(\theta)\sum_n
\delta(t-t_n)\:,
\end{equation}
with the kicking potential
\begin{equation}\label{NHKicking}
V_K(\theta)= K\cos (\theta)+{i} \lambda\cos (\theta+\phi)\;,
\end{equation}
where $p=-i\ehbar\partial/\partial \theta$ is the angular momentum operator, $\theta$ is the angle coordinate, satisfying the commutation relation $[\theta,p]=i\ehbar$ with $\ehbar$ the effective Planck constant. The parameters $K$ and $\lambda$ represent the strength of the real and imaginary components of the kicking potential, respectively. The relative phase  between these two components is determined by the parameter $\phi$. This kind of complex potential has been realized in the atom-optics experiment~\cite{Keller97}. The eigenequation of angular momentum operator is $p|\varphi_n\rangle = p_n |\varphi_n \rangle$ with eigenvalue $p_n = n\ehbar$ and eigenstate $\langle \theta|\varphi_n\rangle=e^{in\theta}/\sqrt{2\pi}$. With the completed basis of $|\varphi_n\rangle$, an arbitrary state can be expanded as $|\psi \rangle=\sum_n \psi_n |\varphi_n\rangle$.
One-period evolution of the quantum state from $t_n$ to $t_{n+1}$ is governed by $|\psi(t_{n+1})\rangle = U|\psi(t_{n})\rangle $, where the Floquet operator $U=U_fU_K$ is composed of the free evolution $U_f =\exp\left(-ip^2/2\ehbar\right)$ and the kicking evolution $U_K =\exp\left[-iV_K(\theta)/\ehbar \right]$.

In the main quantum resonance situation $\ehbar=4\pi$, each matrix elements of $U_f$ is unity, i.e., $U_f(p_n)=\exp\left(-i{n}^2 2\pi\right)=1$, therefore, it has no effects on the time evolution of quantum states. One can get the exact express of the quantum state after arbitrary kick period, i.e., $|\psi(t)\rangle = U_K^{t}|\psi(t_0)\rangle$. Without loss of generality, we choose the ground state of the angular momentum operator as the initial state, i.e., $\psi(\theta,t_0)=1/\sqrt{2\pi}$. Then, the quantum state $|\psi(t)\rangle$ in coordinate space has the expression
\begin{equation}\label{QStateTn}
\psi(\theta,t)=\frac{1}{\sqrt{2\pi}}\exp{\left\{\frac{-it}{4\pi}\left[K\cos (\theta)+{i} \lambda\cos (\theta+\phi)\right]\right\}}\:,
\end{equation}
whose norm takes the form
\begin{equation}\label{QStateNorm}
\mathcal{N}(t)=\int_{-\pi}^{\pi}|\psi(\theta,t)|^2d\theta=I_0\left(\frac{\lambda t}{2\pi}\right)\:.
\end{equation}
Here, $I_0(x)$ denotes the modified Bessel function of the first kind with zeroth order~\cite{Appedix}.
The non-unitary evolution is characterized by the unbounded growth of the $\mathcal{N}(t)$ with time, i.e., $\mathcal{N}(t)\approx \exp\left({\lambda t}/{2\pi}\right)\sqrt{{2\pi}/{\lambda t}}$ for ${\lambda t}/{2\pi}\gg 1$.

In the present work, we investigate both theoretically and numerically the dynamics of the momentum current $\langle p(t)\rangle$, energy diffusion $\langle p^2(t)\rangle$, and quantum scrambling $C(t)=-\langle [A(t),B]^2\rangle$~\cite{ZQi23,XDHu23,JWang21pre}. Note that the $C(t)$ is defined in the Heisenberg picture, with $A(t)=U^{\dagger}(t)AU(t)$ and $\langle \cdot\rangle=\langle \psi(t_0)|\cdot|\psi(t_0)\rangle$ indicates the expectation value of the operator with respect to the initial state~\cite{Pappalardi22,LewisSwan19}. To reduce the impact of the norm to observables, we introduce the rescaled quantities as $\langle p(t)\rangle =\sum_n p_n|\psi_n|^2/\mathcal{N}(t)$ and $\langle p^2(t)\rangle =\sum_n p_n^2|\psi_n|^2/\mathcal{N}(t)$. We use the operators $A=e^{-i\varepsilon p}$ and $B=|\psi(t_0)\rangle\langle\psi(t_0)|$ to construct OTOCs. Straightforward derivation yields the relation $C(t)=\mathcal{N}^2(t)-|\langle \psi(t)|e^{-i\varepsilon p}|\psi(t)\rangle|^2$. Then, a natural definition of the rescaled OTOCs is given by $C(t)=1-|\langle \psi(t)|e^{-i\varepsilon p}|\psi(t)\rangle|^2/\mathcal{N}^2(t)$~\cite{Zhao23arx07}. We consider the case $\varepsilon \ll 1$. Our main results are described by the three following relations
\begin{equation}\label{MMomentlast1}
\langle p(t)\rangle=-K\sin(\phi)\frac{I_1\left(\frac{\lambda t}{2\pi}\right)}{I_0\left(\frac{\lambda t}{2\pi}\right)}t\;,
\end{equation}
\begin{equation}\label{MEnergy5}
	\begin{aligned}
        \langle p^2(t)\rangle
        =&K^2\sin^2(\phi) t^2\\
		&+\frac{2\pi}{\lambda}\frac{I_1\left(\frac{\lambda t}{2\pi}\right)}{I_0\left(\frac{\lambda t}{2\pi}\right)}\left[K^2\cos(2\phi)+\lambda^2\right] t\;,
    \end{aligned}
\end{equation}
and
\begin{equation}\label{FOTCGwh}
\begin{aligned}
C(t)\approx& K^2\varepsilon^2\sin^2(\phi)t^2\left\{1- \left[\frac{I_1\left(\frac{\lambda t}{2\pi}\right)}{I_0\left(\frac{\lambda t}{2\pi}\right)}\right]^2\right\}\\
&+\frac{2\pi\varepsilon^2 t}{\lambda}\frac{I_1\left(\frac{\lambda t}{2\pi}\right)}{I_0\left(\frac{\lambda t}{2\pi}\right)}\left[K^2\cos(2\phi)+\lambda^2\right]\;,
\end{aligned}
\end{equation}
where $I_1(x)$ denotes the modified Bessel function of the first kind with order one~\cite{Appedix}.
\begin{figure}[t]
\begin{center}
\includegraphics[width=8.5cm]{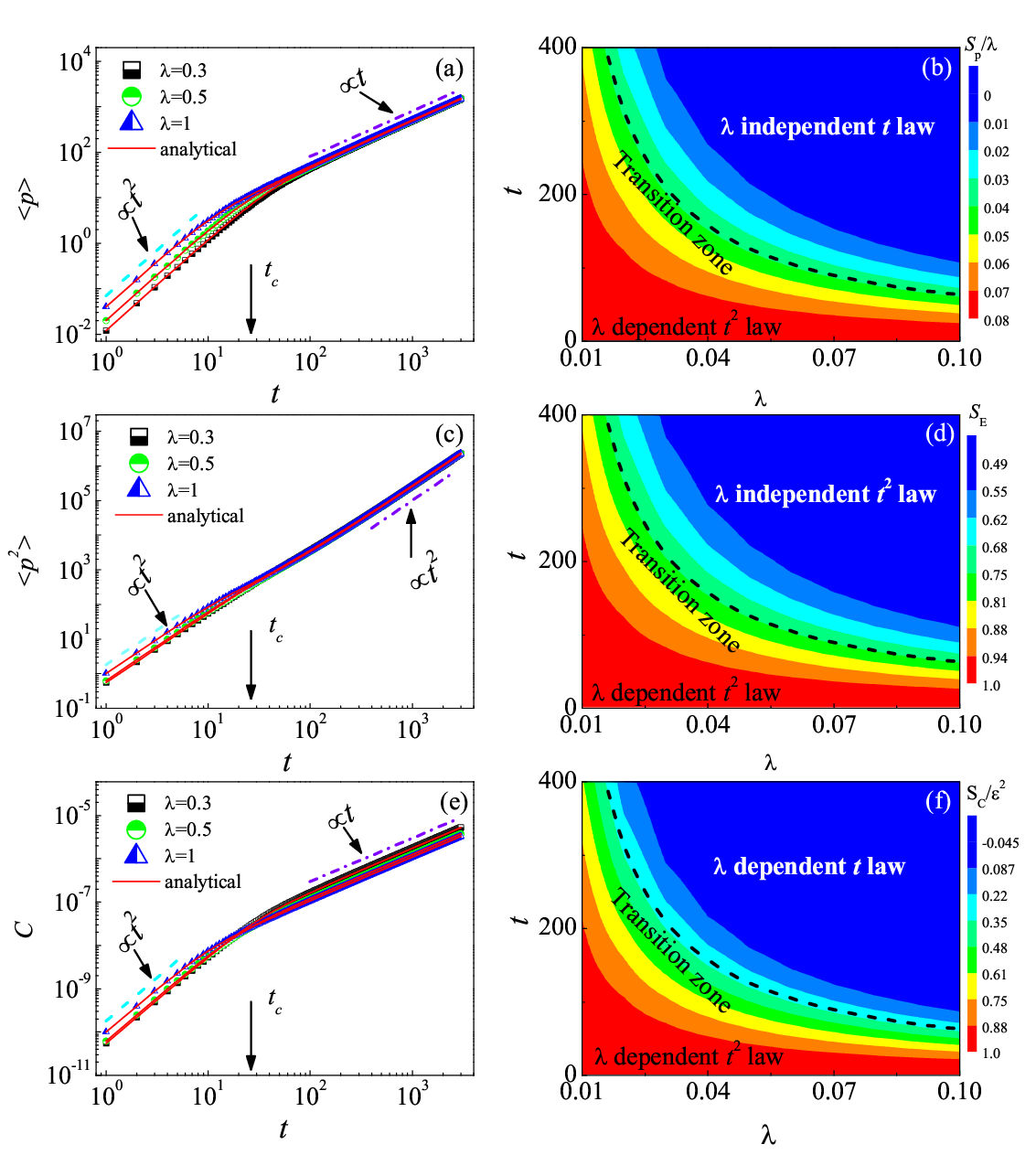}
\caption{Left panels: Time dependence of the $\langle p\rangle$ (a), $\langle p^2\rangle$ (c), and $C$ (e) for $\lambda=0.3$ (squares), 0.5 (circles), and 1 (triangles). Red solid lines in (a), (c), and (f) indicate our theoretical predictions in Eqs.~\eqref{MMomentlast1}, ~\eqref{MEnergy5}, and ~\eqref{FOTCGwh}, respectively. Arrows mark the threshold value of $t_c$. Geen-dashed lines denote a square function of time. Violet dash-dotted lines in (a) and (e) denote the linear function of time, while in (b) it indicates the square function of time. Right panels: The values of $S_p/\lambda$ (b), $S_E$ (d), and $S_C/\varepsilon^2$ (f) in the parameter space $(t,\lambda)$, which show three distinct zones. Dashed lines denote $t_c=2\pi/\lambda$. In (e) and (f), the translation parameter is $\varepsilon=10^{-5}$. The parameters are $K=1$ and $\phi=-\pi/6$.}\label{QCrossover}
\end{center}
\end{figure}

\begin{figure}[t]
\begin{center}
\includegraphics[width=8.0cm]{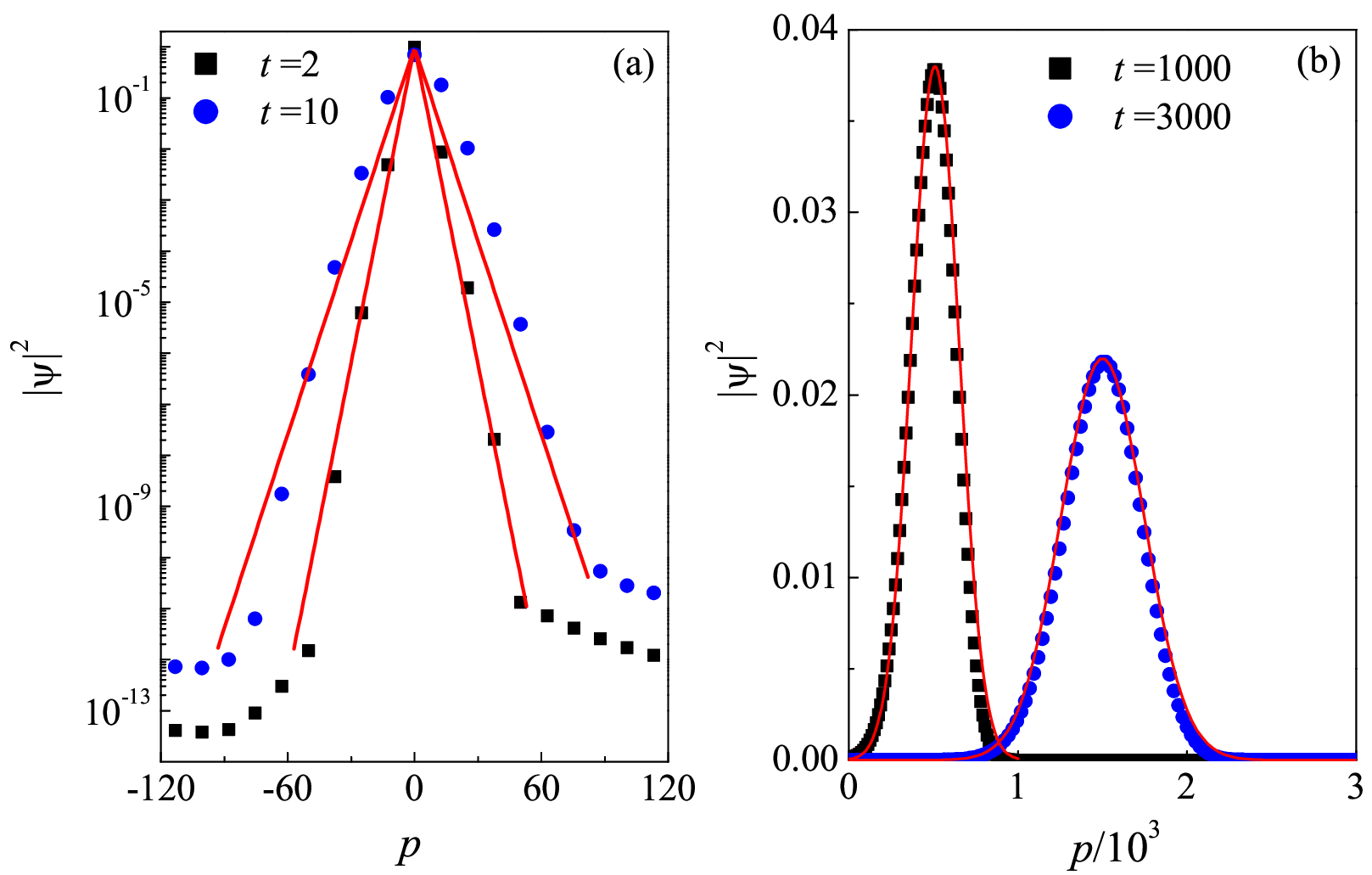}
\caption{Momentum distributions for short (a) and long (b) time evolution with $\lambda=0.3$. In (a), red solid lines indicates the exponential fitting $|\psi(p)|^2\sim \exp(-|p|/\xi)$ with $\xi=2.1$ and 3.43 for $t=2$ and 10 respectively. In (b), red solid lines indicates the Gaussian function fitting $|\psi(p)|^2\sim \exp[-(p-p_c)^2/\sigma]$ with $(p_c=500,\sigma=3.9\times 10^4)$ and $(p_c=1500,\sigma=1.2\times 10^5)$ for $t=1000$ and 3000 respectively. Other parameters are same as in Fig.~\ref{QCrossover}(a).}\label{MDistri}
\end{center}
\end{figure}

\subsection{Directed current}

Figure~\ref{QCrossover}(a) shows that, for a specific $\lambda$ (e.g., $\lambda=0.3$), the $\langle p\rangle$ increases in the quadratic function of time for short time evolutions, and eventually transitions to linear growth. Such a transition occurs around a critical time $t_c$. In addition, one can see perfect agreement between numerical results and theoretical prediction in Eq.~\eqref{MMomentlast1}. For ${\lambda t}/{2\pi}\ll 1$, we have the approximations $I_0\left({\lambda t}/{2\pi}\right)\approx 1$ and $I_1\left({\lambda t}/{2\pi}\right)\approx {\lambda t}/{4\pi}$~\cite{Appedix}. Taking these relations to Eq.~\eqref{MMomentlast1} yields the relation $\langle p(t)\rangle\approx {-K\lambda\sin(\phi)}t^2/{4\pi}$. Apparently, the growth rate $\langle p(t)\rangle/t^2={-K\lambda\sin(\phi)}/{4\pi}$ increases with the increase of $\lambda$, which is validated by our numerical results. For ${\lambda t}/{2\pi}\gg 1$, substituting the approximations of both $I_0\left({\lambda t}/{2\pi}\right)\approx {\exp({\lambda t}/{2\pi})}\left(1+{\pi}/{4\lambda t}\right)/{\sqrt{\lambda t}}$ and $I_1\left({\lambda t}/{2\pi}\right)\approx {\exp({\lambda t}/{2\pi})}\left(1-{3\pi}/{4\lambda t}\right)/{\sqrt{\lambda t}}$~\cite{Appedix} into Eq.~\eqref{MMomentlast1} results in the linear growth $\langle p(t)\rangle\approx -K\sin(\phi)(t-\pi /|\lambda|)$, which is independent on $\lambda$ when $t\gg 1$. The above estimations uncover the mechanism for the transition from quadratic-law growth to the linear growth as time evolves. In addition, our findings of the  sinusoidal relationship between mean momentum and phase $\phi$ pave the way for Floquet engineering of the directed current in non-Hermitian chaotic systems.

To reveal whether there is singularity for the transition of the $\langle p \rangle$ from the short time to long time behavior, we investigate the second derivative of the mean momentum $S_p= {d^{2}\langle p(t)\rangle}/{dt^2}$. Note that, in the following derivation, for brevity, we use $I_j$ to replace $I_j(\frac{\lambda t}{2\pi})$ ($j=0,1,2...$). The analytical expression takes the form
\begin{equation}\label{D2Current}
\begin{aligned}
S_p=&-K\sin(\phi)\frac{\lambda}{2\pi}\left[1+\frac{I_2}{I_0}-2\left(\frac{I_1}{I_0}\right)^2\right]\\
&+K\sin(\phi)\frac{\lambda^{2}t}{4\pi^{2}}\left[\frac{3I_1-I_3}{4I_0}+\frac{3I_1I_2}{2I^{2}_0}-2\left(\frac{I_1}{I_0}\right)^{3}\right]\;.
\end{aligned}
\end{equation}
Taking into account the approximation of $I_j$ ($j=0,1,2,3$) under two different conditions, namely, when $\lambda t/2\pi \ll 1$ and when $\lambda t/2\pi \gg 1$~\cite{Appedix}, we can derive approximate expressions
\begin{align}\label{SDrivMEAM}
S_p \approx
\begin{cases}
\frac{-K\sin(\phi)\lambda}{2\pi}\;, & \text{for $\frac{\lambda t}{2\pi} \ll 1$}\;,\\
0\;,  & \text{for $\frac{\lambda t}{2\pi} \gg 1$}\;,
\end{cases}
\end{align}
In Fig.~\ref{QCrossover}(b), we have plotted the ratio $S_p/\lambda$ for various values of $t$ and $\lambda$. Three distinct zones can be observed: i) a $\lambda$ dependent $t^2$-law zone with $S_p/\lambda=-K\sin(\phi)/2\pi$ for $t\ll t_c=2\pi/\lambda$; ii) a $\lambda$ independent $t$-law zone with $S_p=0$ for $t\gg t_c$; and iii) a transition zone for $t\sim t_c$.

In the $\lambda$ dependent $t^2$-law zone, the quantum state is exponentially localized in momentum space, i.e., $|\psi(p)|^2\sim \exp(-|p|/\xi)$, whose localization length $\xi$ increases with time [see Fig.~\ref{MDistri}(a)]. Detailed observations reveal that this exponentially-localized shape of the quantum state is asymmetric around $p=0$. This kind of asymmetric spreading of the quantum state in momentum space results in the growth of mean momentum with time. While in the $\lambda$ independent $t$-law zone, the momentum distribution can be well described by the Gaussian function $|\psi(p)|^2\sim \exp[-(p-p_c)^2/\sigma]$ [see Fig.~\ref{MDistri}(b)]. Interestingly, the comparison of the momentum distribution in different time demonstrates that the center momentum $p_c$ linearly increases with time, i.e., $p_c(t)=Dt$ for which the growth rate equal to that of the $\langle p(t)\rangle$, i.e., $D=d\langle p(t)\rangle/dt$. Therefore, the linear growth of the mean momentum for $t\gg t_c$ originates from the directed movement of the soliton-like wavepacket in momentum space. In addition, we would like to stress that the width $\sigma$ of the Gaussian wavepacket also increases with time.

\subsection{Energy diffusion}

Figure~\ref{QCrossover}(c) illustrates that the NQKR exhibits ballistic energy diffusion over time, specifically, $\langle p^2(t)\rangle = G t^2$, which is in perfect agreement with our theoretical prediction in Eq.~\eqref{MEnergy5}. Interestingly, for $t\ll t_c$, the diffusion rate $G$ increases with the non-Hermitian parameter $\lambda$, and for $t\gg t_c$, it becomes independent of $\lambda$. By approximating both $I_0(\lambda t/2\pi)$ and $I_1(\lambda t/2\pi)$ under two different limits, we can derive the approximate expression for Eq.~\eqref{MEnergy5}, namely, $\langle p^2(t)\rangle \approx (K^2 + \lambda^2)t^2/2$ for $t \ll t_c$ and $\langle p^2(t)\rangle \approx K^2\sin^2(\phi)t^2 +{2\pi t}\left[K^2\cos(2\phi)+\lambda^2\right]/{|\lambda|}$ for $t \gg t_c$. This confirms the transition from $\lambda$-dependent behavior to $\lambda$-independent behavior. This transition can also be observed in the second derivative of the mean square of momentum, denoted as $S_E = d^2 \langle p^2(t)\rangle/d t^2$,
\begin{equation}\label{D2Energy}
\begin{aligned}
S_E=2K^2\sin^2(\phi) - \frac{2\pi S_p}{K\lambda \sin(\phi)}\left[K^2\cos(2\phi)+\lambda^2\right]\;,
\end{aligned}
\end{equation}
which can be approximated as
\begin{align}\label{SDrivMEAE}
S_E \approx
\begin{cases}
K^2 + \lambda^2\;, & \text{for $\frac{\lambda t}{2\pi} \ll 1$}\;,\\
2K^2\sin^2(\phi)\;,  & \text{for $\frac{\lambda t}{2\pi} \gg 1$}\;.
\end{cases}
\end{align}
In Figure~\ref{QCrossover}(d), we present numerical results of $S_E$ based on Eq.~\eqref{D2Energy}, which demonstrates the $\lambda$-dependent zone for $t\ll t_c$, a transition zone for $t \sim t_c$, and the $\lambda$-independent zone for $t\gg t_c$. Our discovery of the sinusoidal dependence of $S_E$ on phase $\phi$ provides a theoretical foundation for engineering the energy diffusion through non-Hermitian driven potential.

\subsection{Quantum scrambling}

Based on the Taylor expansion $e^{-i\varepsilon p}\approx 1-i\varepsilon p$ for $\varepsilon \ll 1$, we obtain the relations
\begin{equation}\label{FOTCGwh1}
\begin{aligned}
C(t)&\approx \varepsilon^2 \left[\langle p^2\rangle -\langle p\rangle^2\right]\;\\
    & = K^2\varepsilon^2\sin^2(\phi)t^2\left[1- \left(\frac{I_1}{I_0}\right)^2\right]\;\\
	&+\frac{2\pi\varepsilon^2 t}{\lambda}\frac{I_1}{I_0}\left[K^2\cos(2\phi)+\lambda^2\right]\;,
\end{aligned}
\end{equation}
which has two different asymptotic behaviors
\begin{align}\label{FOTCLGwh}
C(t)\approx
\begin{cases}
\frac{\varepsilon^2\left(K^2+\lambda^2\right)}{2}t^2\;, & \text{for $\lambda t /2\pi\ll 1$}\;,\\
\frac{2\pi\varepsilon^{2}}{|\lambda|}\left[\frac{1+\cos(2\phi)}{2}K^2+\lambda^2\right]t\;,  & \text{for $\lambda t /2\pi \gg 1$}\;.
\end{cases}
\end{align}
In Fig.~\ref{QCrossover}(c), we show our numerical results for the time evolution of $C$ for different $\lambda$. It is evident that $C$ exhibits a quadratic growth with time for $t\ll t_c$ and a linear growth with time for $t \gg t_c$, with both behaviors being dependent on $\lambda$. These distinct behaviors are also characterized by the second derivative $S_C= d^2 C(t)/dt^2$. Straightforward derivation yields the relation
\begin{equation}\label{D2OTOC}
S_C= \varepsilon^2 \left[S_E -2 \left(\frac{d\langle p\rangle}{dt}\right)^2-2\langle p\rangle S_p \right]\;,
\end{equation}
with
\begin{equation}\label{DCurrent}
\frac{d\langle p\rangle}{dt}=-K\sin(\phi)\left\{\frac{\lambda t}{4\pi}\left[1+\frac{I_2}{I_0}-2\left(\frac{I_1}{I_0}\right)^2\right]+\frac{I_1}{I_0}\right\}\;.
\end{equation}
By approximating the terms on the right side of Eq.\eqref{D2OTOC} under two different limits, we derive the approximate relation
\begin{align}\label{SDOTCAppro}
S_C\approx
\begin{cases}
\varepsilon^2\left(K^2+\lambda^2\right)\;, & \text{for $\lambda t/2\pi \ll 1$}\;,\\
0\;,  & \text{for $\lambda t/2\pi \gg 1$}\;.
\end{cases}
\end{align}
Our numerical results, based on Eq.~\eqref{D2OTOC}, clearly demonstrate the
$\lambda$-dependent quadratic growth zone for $t\ll t_c$, the transition zone for $t \sim t_c$, and the $\lambda$-dependent linear growth zone for $t\gg t_c$ [see Fig.~\ref{QCrossover}(f)], validating our theoretical prediction in Eq.\eqref{SDOTCAppro}.

\subsection{Some remarks on the relation between phase manipulation and $\cal{PT}$-symmetry}
\begin{table}[b]
\begin{center}
\begin{tabular}{|m{3.7cm}<{\centering}|m{1.5cm}<{\centering}|m{3.0cm}<{\centering}|}
\hline
Phase $\phi$ & $\pi/2$ & $\pi$\\\hline
Symmetry class & $\mathcal{PT}$ & non-$\mathcal{PT}$ \\\hline
Directed current $\langle p(t) \rangle\propto K\sin(\phi)t$& $\propto Kt$ & $0$\\\hline
Energy diffusion $\langle p^2(t)\rangle\propto K^2\sin^2(\phi)t^2 +{2\pi }\left[K^2\cos(2\phi)+\lambda^2\right]t/{|\lambda|}$ & $\propto K^2t^2$ & $\propto \frac{2\pi }{|\lambda|}\left(K^2+\lambda^2\right)t$\\\hline
Quantum scrambling $C(t)\propto\frac{2\pi\varepsilon^{2}}{|\lambda|}\left[\frac{1+\cos(2\phi)}{2}K^2+\lambda^2\right]t$ & $\propto {2\pi\varepsilon^{2}}\lambda t$ & $\propto \frac{2\pi\varepsilon^{2}}{|\lambda|}\left(K^2+\lambda^2\right)t$\\\hline
\end{tabular}
\caption{Symmetry and the long-time behavior of the $\langle p \rangle$, $\langle p^2\rangle$, and $C$ for $\phi=\pi/2$ and $\pi$.}\label{ConcTabl}
\end{center}
\end{table}

The dependence of the long-time behavior of the $\langle p \rangle$, $\langle p^2\rangle$ and $C$ on the phase $\phi$ opens the opportunity for the manipulation of both the quantum transport and quantum scrambling via the relative phase between the real part and the imaginary part of the kicking potential $V_K(\theta)= K\cos (\theta)+{i} \lambda\cos (\theta+\phi)$ [see Eq.~\eqref{NHKicking}].
Interestingly, the  potential is $\mathcal{PT}$-symmetric when $\phi=\pi/2$. In this situation, both directed current and energy diffusion reach their maximum values, namely, $\langle p \rangle\approx -Kt$ and $\langle p^2\rangle \approx K^2t^2$ since $\langle p \rangle$ and $\langle p^2 \rangle$ are sinusoidal functions of $\phi$, signaling the $\cal{PT}$-symmetry protected transport behaviors. In contrast, quantum scrambling is minimized, i.e., $C\approx 2\pi \varepsilon^2 |\lambda| t$ because $C$ behaves as a cosine function of $2\phi$ (see Table.~\ref{ConcTabl}). For $\phi=\pi$, the NQKR model is a general non-Hermitian system that does not have $\cal{PT}$-symmetry, for which the directed current is totally suppressed, namely, $\langle p \rangle=0$, and the energy diffusion reduces as $\langle p^2\rangle \propto t$. In contrast the $C$ is maximum $C\approx {2\pi \varepsilon^{2}t}(K^2+\lambda^2)/{|\lambda|}$ demonstrating the non-Hermiticity enhanced quantum scrambling.

\section{Conclusion and discussions}\label{SecSum}

One effective strategy for modulating quantum dynamics involves implementing Floquet driven potentials in systems, which has been accomplished by state-of-the-art experiments in both atom optics and optical waveguides. In this work, we investigate the interesting problems of the phase modulation of the directed current, energy diffusion and quantum scrambling, in a NQKR model with quantum resonance condition. We uncover a dynamical crossover in time-dependent behaviors of these phenomena. For short time interval $t\ll t_c$, the $\langle p \rangle$, $\langle p^2\rangle$, and $C$ all exhibit quadratic growth with time, and their growth rates depend on $\lambda$. After sufficiently long time evolution $t\gg t_c$, the $\langle p \rangle$ shows a linear growth independent of $\lambda$, the $\langle p^2\rangle$ tansitions to the $\lambda$-independent quadratic growth with time, and the $C$ exhibits a linear growth dependent on the $\lambda$. These exotic behaviors are in perfect agreement with our theoretically predictions. Based on the second derivative, namely, $S_p$, $S_E$ and $S_C$ of these observables, we obtain the phase diagram of the crossover in the parameter space $(t, \lambda)$, which clearly presents three distinct zones for the dynamics of the momentum current, quantum diffusion and quantum scrambling.

In the atom-optics experiment, the complex potential in Eq.\eqref{NHKicking} has been realized by the superposition of two standing laser fields, interacting with ultracold atoms that encompass a ground state $E_1$, excited states with two hyperfine levels $E_2^{\pm}$, and a non-interacting state $E_i$\cite{Keller97,Chudesnikov91}. The far-tuned standing laser, coupling $E_1$ and $E_2^{-}$, generates a dipole force on the atoms, emulating the real component of the complex potential. The resonance laser facilitates the transition from $E_1$ to $E_2^{+}$. The ultracold atoms in $E_2^{+}$ subsequently transition to $E_i$, resulting in particle loss and thereby mimicking the imaginary part of the complex potential. The precise modulation of the relative phase between the real and imaginary parts of the complex potential is achievable by adjusting the distance between the atomic beam and the mirror surface. The delta-kicking potential is emulated using a sequence of short square pulses created through the modulation of the standing wave by an acousto-optical modulator~\cite{Lopez13}. Consequently, our system in Eq.~\eqref{Hamil} is achievable in the state-of-the-art atom-optics experiments. Both the expectation values, such as $\langle p\rangle$ and $\langle p^2\rangle$, and the variance can be detected via time-of-flight measurement of the probability density in momentum space~\cite{Hainaut18prl,Hainaut18nc}, which paves the way for the experimental validation of our findings.

\section*{Acknowledgments}
Wen-Lei Zhao is supported by the National Natural Science Foundation of China (Grant Nos. 12065009 and 12365002), the Natural Science Foundation of Jiangxi province (Grant Nos. 20224ACB201006 and 20224BAB201023) and the Science and Technology Planning Project of Ganzhou City (Grant No. 202101095077). Jie Liu is supported by the NSAF (Contract No. U2330401).

\appendix
\section{Properties of the modified Bessel functions of the first kind}
The modified Bessel functions of the first kind are defined by
\begin{equation}\label{MDFBessel}
I_m\left(x\right)=\frac{1}{2\pi}\int_0^{2\pi}e^{-im\theta}\exp\left[x\cos(\theta)\right]d\theta\;.
\end{equation}
The function $I_m(x)$ can be expanded as
\begin{widetext}
\begin{equation}\label{ExpanMFBes}
I_m(x)=\frac{e^{x}}{\sqrt{2\pi x}}\left[1- \frac{4m^2-1}{8x}+\frac{(4m^2-1)(4m^2-9)}{2!(8x)^2}-\frac{(4m^2-1)(4m^2-9)(4m^2-25)}{3!(8x)^3}+\cdots\right]\;.
\end{equation}
\end{widetext}
Our theoretical analysis involves the function $I_m(x)$ with $0 \leq m \leq 3$. For $x\ll 1$, they can be approximated as
\begin{equation}\label{SmallApp}
\begin{aligned}
I_0\left(x\right)\approx 1\;,I_1\left(x\right)\approx
\frac{x}{2}\;,I_2\left(x\right)\approx
\frac{x^2}{8}\;, \text{and}\;
I_3\left(x\right)\approx\frac{x^3}{48}\;.
\end{aligned}
\end{equation}
For $x\gg 1$, the two leading terms in Eq.~\eqref{ExpanMFBes} contribute significantly. Therefore, we have neglected the other terms and obtained the following relations
\begin{equation}\label{LargeApp0}
I_0\left(x\right)\approx\frac{e^{x}}{\sqrt{2\pi x}}\left(1+\frac{1}{8x}\right)\;,
\end{equation}
\begin{equation}\label{LargeApp1}
I_1\left(x\right)\approx\frac{e^{x}}{\sqrt{2\pi x}}\left(1- \frac{3}{8x}\right)\;,
\end{equation}
\begin{equation}\label{LargeApp2}
I_2\left(x\right)\approx\frac{e^{x}}{\sqrt{2\pi x}}\left(1-\frac{15}{8x}\right)\;,
\end{equation}
and
\begin{equation}\label{LargeApp3}
I_3\left(x\right)\approx\frac{e^{x}}{\sqrt{2\pi x}}\left(1- \frac{35}{8x}\right)\;.
\end{equation}

\end{document}